\documentclass[fleqn,usenatbib]{mnras}

\usepackage{newtxtext,newtxmath}

\usepackage[T1]{fontenc}
\usepackage{ae,aecompl}

\usepackage{graphicx}	
\usepackage{amsmath}	
\usepackage{amssymb}	
\usepackage{wasysym}
\usepackage{threeparttable}
\usepackage[dvipsnames]{xcolor}

\title[Planetary Magnetism as a Parameter in Exoplanet Habitability]{Planetary Magnetism as a Parameter in Exoplanet Habitability}

\author[S. R. N. McIntyre et al.]{
Sarah R.N. McIntyre,$^{1}$\thanks{E-mail: sarah.mcintyre@anu.edu.au}
Charles H. Lineweaver$^{1,2}$ and
Michael J. Ireland$^{1}$
\\
$^{1}$Research School of Astronomy and Astrophysics, Australian National University, Canberra, ACT 2611, Australia\\
$^{2}$Research School of Earth Sciences, Australian National University, Canberra, ACT 2611, Australia\\
}

\date{Accepted XXX. Received YYY; in original form ZZZ}

\pubyear{2019}

\begin{document}
\label{firstpage}
\pagerange{\pageref{firstpage}--\pageref{lastpage}}
\maketitle

\begin{abstract}
Evidence from the solar system suggests that, unlike Venus and Mars, the presence of a strong magnetic dipole moment on Earth has helped maintain liquid water on its surface. Therefore, planetary magnetism could have a significant effect on the long-term maintenance of atmosphere and liquid water on rocky exoplanets. We use Olson \& Christensen's (2006) model to estimate magnetic dipole moments of rocky exoplanets with radii R${}_{p}$ $\le$ 1.23 R$_\oplus$. Even when modelling maximum magnetic dipole moments, only Kepler-186 f has a magnetic dipole moment larger than the Earth's, while approximately half of rocky exoplanets detected in the circumstellar habitable zone have a negligible magnetic dipole moment. This suggests that planetary magnetism is an important factor when prioritizing observations of potentially habitable planets.
\end{abstract}




\section{Introduction}

Within the next decade, upcoming observations with near-future telescopes will detect an ever-increasing number of exo-Earths. In order to make the most of the limited observational resources available, target selection has focused on `habitable worlds' defined as rocky bodies (with enough surface gravity to sustain an atmosphere) orbiting their host stars at a distance where stellar radiation is suitable for the presence of surface liquid water \citep{kaltenegger2017characterize}. However, numerous planetary and astronomical factors influence an exoplanet's ability to maintain liquid water. It is increasingly important to expand our considerations to multiple parameters including magnetic field, albedo, stellar type, planet chemical composition, orbital eccentricity, inclination, tidal locking, impact events, and plate tectonics. This will enable us to prioritize planets most likely to maintain liquid water in order to best utilize telescope time when biosignature observations become a possibility. 

Most assessments of habitability stem from observations of our own solar system. The presence of a magnetic dipole moment on Earth protects the surface and liquid water from solar winds and flares \citep{elkins2013makes}. Venus, Earth and Mars likely began with similar amounts of water \citep{grinspoon1993implications,way2016venus}. This is corroborated by their deuterium to hydrogen ratios (D/H) which suggest that both Mars and Venus had more water early in their histories and have lost most of it. Venus' atmosphere has a D/H abundance ratio $\sim$100 D/H of bulk Earth \citep{lammer2008atmospheric}. Without the protection of a magnetic field, water vapour photodissociated in the upper atmosphere due to enhanced early solar EUV irradiation \citep{kasting1983loss}. This led to the escape of hydrogen into space and resulted in rapid surface liquid water loss \citep{lammer2011hydrogen}. In 2015 the Maven mission confirmed that water was abundant and active on Mars for the first few hundred million years of the solar system \citep{jakosky2015mars}. However, at some point less than a billion years after Mars formed, its global magnetic field went extinct, removing a source of protection from the solar winds \citep{stevenson2001mars,golombek2010mars}. The loss of water raised Mars' atmospheric D/H ratio to its current value $\sim$5 D/H of Earth \citep{hallis2017d}.

Almost all of the currently detected rocky exoplanets are on close, highly-irradiated orbits, bombarded by large amounts of ionizing EUV and X-ray radiation \citep{owen2012planetary}. With weak or negligible magnetic protection, the upper atmosphere of an exoplanet will be more exposed to stellar winds and coronal mass ejections, resulting in atmospheric mass-loss due to non-thermal processes such as ion pickup, photo-chemical energizing mechanisms, and sputtering \citep{vidotto2013effects}. Furthermore, even in the absence of stellar wind effects, the closed magnetic field lines above an exosphere inhibit the loss of charged particles to interplanetary space \citep{lammer2002martian}. Since planetary magnetism reduces non-thermal atmospheric erosion, it affects the evolution of a planet's environment and its potential habitability \citep{gudel2014astrophysical}. Here we focus on modelling the strengths of the magnetic dipole moments of rocky exoplanets to assess their ability to protect their atmospheres from hydrogen loss ($\sim$ water loss). 
\section{Method}

\subsection{Magnetic moment model description}
\label{sec:magmod}

In our calculations we use the \citet{olson2006dipole} magnetic moment scaling law:

\begin{equation}
\label{eq:magmom}
    \mathcal{M}=4\pi {{\mathrm{r}}_0}^3\gamma {\left(\frac{{\overline{\rho }}_0}{{\mu }_0}\right)}^{1/2}{\left(FD\right)}^{1/3}	
\end{equation}

\noindent where $\mathcal{M}$ is the magnetic moment (in {A m}${}^{2}$), r${}_{0}$ is the planet's core radius, a fitting coefficient $\gamma = 0.15\pm 0.05$ is inferred from numerical simulations \citep{olson2006dipole}, ${\overline{\rho }}_0$ is the bulk density of the outer liquid core (we use the Earth's density model: ${\overline{\rho }}_0=11000\pm 1100\ \textnormal{kg}\ \mathrm{m}^{-3}$ \citep{litasov2016composition}), the magnetic permeability of the vacuum ${\mu }_0=4\pi \times {10}^{-7}\mathrm{H/m}$, \textit{F} is the average convective buoyancy flux, and \textit{D} is the thickness of the outer liquid core.  

One of the factors needed for equation (\ref{eq:magmom}) is planetary core radius, r${}_{0}$. \citet{zeng2016mass} used direct extrapolations from Earth's seismic model to integrate pressure and density along adiabatic profiles. These integrations provide radial density profiles for rocky exoplanets with an iron core, from which \citet{zeng2016mass} proposed a semi-empirical relationship between core mass fraction (CMF), planetary radius, and mass. \citet{zeng2017simple} then used the assumption that the internal gravity profile can be approximated as a piecewise function to show that the CMF can be related to the core radius fraction (CRF) of a rocky planet as $\mathrm{CRF}\mathrm{\approx }\sqrt{\mathrm{CMF}}$. The planetary core radius can then be calculated as: 

\begin{equation} 
\label{eq:corad} 
\frac{{\mathrm{r}}_0}{{\mathrm{R}}_{\mathrm{p}}}\mathrm{=}\sqrt{\frac{\mathrm{1}}{\mathrm{0.21}}\left[\mathrm{1.07-}\left(\frac{{\mathrm{R}}_{\mathrm{p}}}{{\mathrm{R}}_{\mathrm{\oplus }}}\right)\mathrm{/}{\left(\frac{{\mathrm{M}}_{\mathrm{p}}}{{\mathrm{M}}_{\mathrm{\oplus }}}\right)}^{\mathrm{0.27}}\right]} 
\end{equation} 

\noindent where R${}_{p}$ and M${}_{p}$ are the planet's radius and mass respectively. In equation~(\ref{eq:magmom}) \textit{F} represents the strength of the thermal and chemical convection-driven dynamo in the planet's outer liquid core \citep{lopez2011magnetic}. \textit{F} is expressed in terms of the local Rossby number $R_{O_l}$, the thickness of the outer liquid core \textit{D}, and the rotation rate \textit{$\mathit{\Omega}$}, with all three normalized to their corresponding Earth values:

\begin{equation} 
\label{eq:dynamo} 
\frac{F}{F_{\oplus }}={\left(\frac{R_{O_l}}{R_{O_{l\oplus }}}\right)}^2{\left(\frac{D}{D_{\oplus }}\right)}^{2/3}{\left(\frac{\mathrm{\Omega }}{{\mathrm{\Omega }}_{\oplus }}\right)}^{7/3} 
\end{equation} 

\noindent where $D=0.65{\mathrm{r}}_0$ is the radial extent of the convection cell based on the \citet{heimpel2005simulation} model for the most efficient dynamo for at least some period of the planet's lifetime. We use a $D=0.65{\mathrm{r}}_0$ conversion for all the rocky exoplanets in our sample and $D_{\oplus} =0.65{\mathrm{r}}_{0_{\oplus}}$ for the outer liquid core radius of the Earth.  \citet{olson2006dipole} calculated a value of 0.09 for Earth's local Rossby number and determined that $R_{O_l}$ must be $\le 0.12$ for a base-heated dynamo to produce a dipolar magnetic moment. Thus, in order to model the best-case scenario, we have set $R_{O_l} = 0.12$ to generate an optimal convective buoyancy flux, allowing us to estimate the \textit{maximum} magnetic dipole moment (${\mathcal{M}}_{max}$).

\subsection{Rotation rate calculations}
\label{sec:rotrat}

Our estimates of rotation rate $\Omega$ depend on whether an exoplanet is tidally locked or non-tidally locked. For the subset of tidally locked exoplanets we assume that the rotation period equals the orbital period. \citet{griessmeier2009protection} was used to calculate the time taken to tidally lock an exoplanet:

\begin{equation} 
\label{eq:rotp} 
{\tau} _{sync}\approx \frac{4}{9}\alpha Q'_p\left(\frac{{R_p}^3}{GM_p}\right)\left({\mathit{\Omega}}_i-{\mathit{\Omega}}_f\right){\left(\frac{M_p}{M_ *}\right)}^2{\left(\frac{d}{R_p}\right)}^6 
\end{equation}

\noindent
where d is the planet's semi-major axis and the structure parameter $\alpha$ is set at Earth's value of $\alpha$ = 1/3. We adopt tidal dissipation factor of $Q'_p = 500$ corresponding to ``small super-Earths'' and ``ocean planets'' \citep{griessmeier2009protection}. The initial and final rotation rates $\Omega_i$ and $\Omega_f$ of a rocky exoplanet are not well-known quantities \citep{correia2003different}. Following our strategy to model the  maximum dipole magnetic moment, we (i) use a high initial rotation rate analogous to early Earth's rotation period of 4 hr prior to the moon-forming impact, corresponding to $\Omega_i = 5.86\Omega_\oplus$ \citep{canup2008accretion}, and (ii) assume $\Omega_f = 0$ \citep{griessmeier2006aspects}. We compare these ${\tau }_{sync}$  values with the stellar age of the host stars (for stars with no age estimates, we assume a lower age limit of 1 Gyr) (Table~\ref{tab:exodat}) and find that $\sim$99\% of rocky exoplanets in our sample are tidally locked. (${\tau} _{sync}$ < stellar age). For these we set $\Omega= \frac{2\pi}{orbital\ period}$ in equation~(\ref{eq:dynamo}). 

\begin{figure}
	\includegraphics[width=\columnwidth]{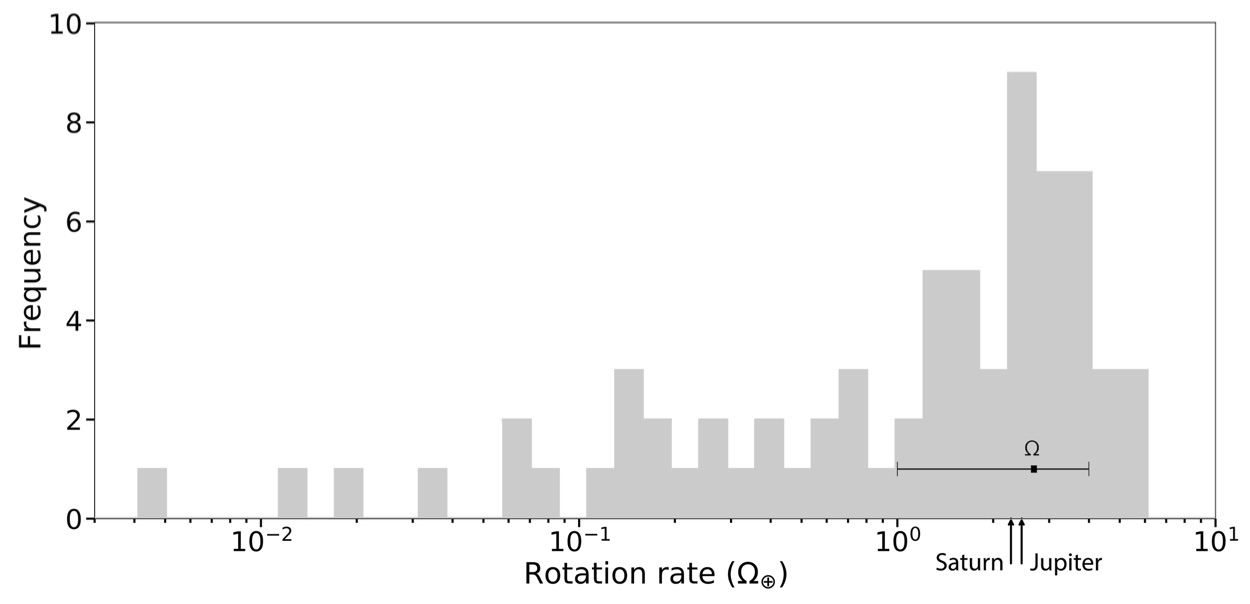}
    \caption{Histogram of the rotation rates of 43 non-tidally locked solar system objects in units of Earth's rotation rate $\Omega_\oplus$ (Table~\ref{tab:rotcal}). The rotation rates of Jupiter and Saturn indicate that a mass-weighted distribution of the 43 solar system objects would result in a similar mean with a narrower spread. We conservatively use this broader, number-weighted distribution.}
    \label{fig:example_rotation-rates}
\end{figure}

Since we do not have the planetary rotation rates for non-tidally locked exoplanets, we examine the rotation rate of 43 solar system objects including planets, moons, dwarf planets, Kuiper belt objects and main-belt asteroids (Fig.~\ref{fig:example_rotation-rates}, Table~\ref{tab:rotcal}). Their distribution is characterised by an average rotation rate of $\Omega=2.5\pm 1.5 \ \Omega_\oplus$ and we assign this value to the five exoplanets in our sample that are not tidally locked. This distribution spans a very broad range, including the low rotational velocities of the few terrestrial planets in our solar system.

\subsection{Sample Selection and Monte Carlo Simulation}
\label{sec:samsel}

Before applying our magnetic moment model, we compile a database of rocky exoplanets using the NASA composite exoplanet database\footnote{\url{https://exoplanetarchive.ipac.caltech.edu/cgi-bin/TblView/nph-tblView?app=ExoTbls&config=compositepars} (accessed 4 January 2019) \label{nasa}}, in conjunction with the higher precision radii of Kepler planets given in \citet{fulton2018california}(Table~\ref{tab:exodat}). Unknown masses or radii and their associated uncertainties are estimated using \citet{chen2016probabilistic} M-R relationship for terrestrial planets: 

\begin{equation} 
\label{eq:masseq} 
{{\mathrm{R}}_{\mathrm{p}}} \sim {{\mathrm{M}}_{\mathrm{p}}}^{\mathrm{0.279 \pm 0.009}} 
\end{equation}

Following the \citet{chen2016probabilistic} definition of the boundary between terrestrial and Jovian planets, we restrict our sample to exoplanets with radii R${}_{p}$ $\le$ 1.23 R$_\oplus$, which corresponds to $\sim$ 2 M$_\oplus$ and agrees with the \citet{gaidos2010thermodynamic} prediction that rocky planets with mass > 2.5 M$_\oplus$ will not develop inner cores and hence will not be able to sustain a magnetic field. These constraints restrict our sample to 496 rocky exoplanets (490 tidally locked), 9 of which are located in the circumstellar habitable zone (CHZ) - where the CHZ is the insolation between recent Venus and ancient Mars as defined by the \citet{kopparapu2013habitable} optimistic habitable zone \citep{kopparapu2013habitable,kopparapu2014habitable}. Eight of the nine rocky exoplanets from our CHZ subset are tidally locked.

For each exoplanet we perform 10,000 Monte Carlo simulations, taking into account uncertainties on the masses, radii, and rotation rates of the exoplanets as given in \citet{fulton2018california} and the NASA composite exoplanet database\textsuperscript{\ref{nasa}}, as well as uncertainties on the parameters $\gamma$ and ${\overline{\rho }}_0$. Furthermore, for the subset of planets where the mass-radius relation was used, equation (\ref{eq:masseq}) was input directly into the Monte Carlo simulation to ensure the uncertainties were appropriately correlated. These simulations allow us to determine the median and 68\% confidence intervals on ${\mathcal{M}}_{max}$ values.

\section{Data Analysis}

\begin{figure}
	\includegraphics[width=\columnwidth]{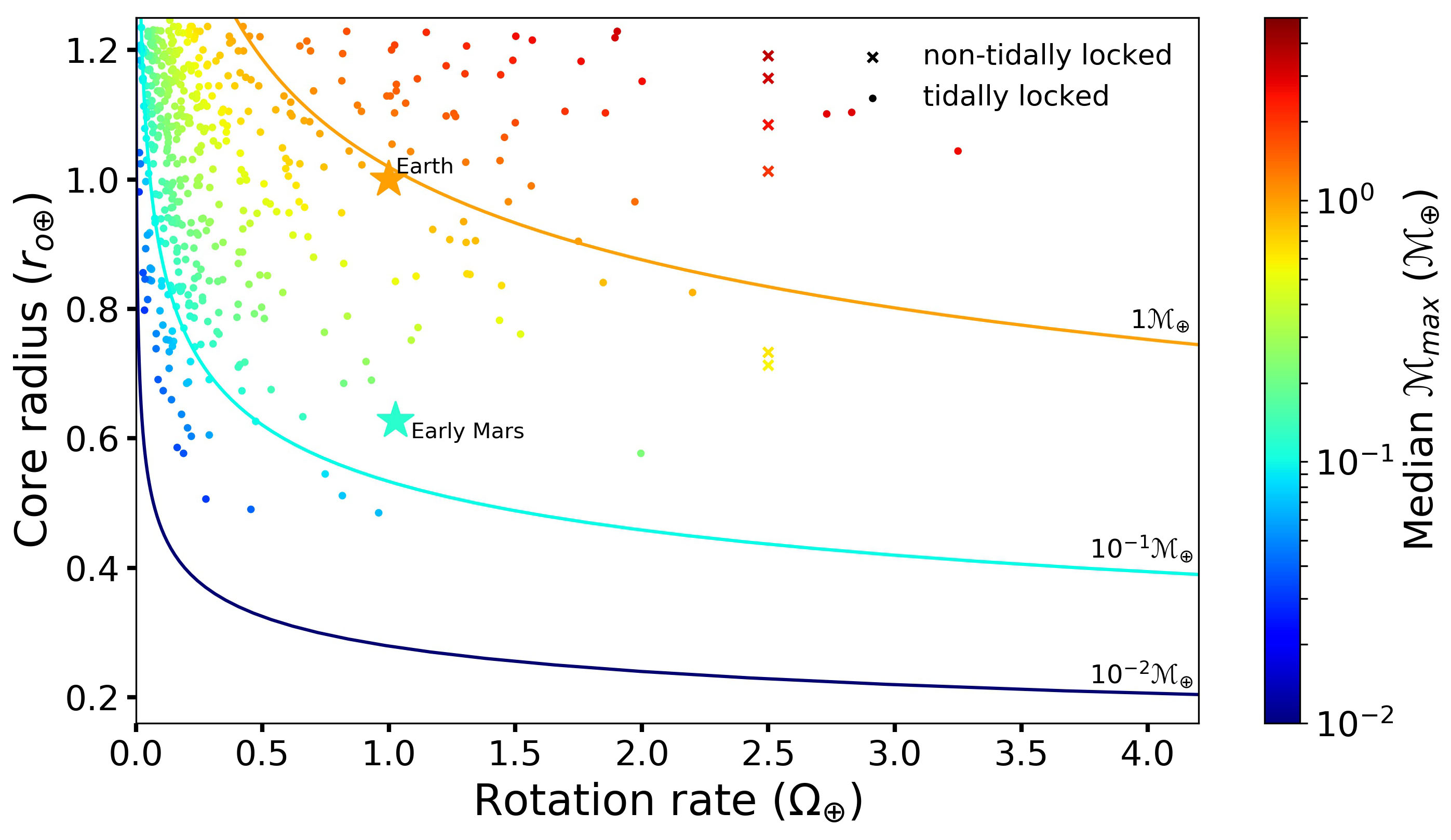}
    \caption{Median ${\mathcal{M}}_{max}$ of 496 exoplanets in our sample as a function of planetary core radius and rotation rate. Error bars on individual exoplanets are omitted for clarity. Non-tidally locked exoplanets have rotation rates based on Fig.~\ref{fig:example_rotation-rates}. The coloured lines are contours of constant ${\mathcal{M}}$ where ${\mathcal{M}} \sim {{\mathrm{r}}_{\mathrm{0}}}^{\mathrm{4}} \Omega^{7/9}$ (see equations~\ref{eq:magmom} - \ref{eq:dynamo}). Since 4 > ${7/9}$ ${\mathcal{M}}_{max}$ is more sensitive to core radius than rotation rate.} 
    \label{fig:example_maxmm}
\end{figure}

The median ${\mathcal{M}}_{max}$ for all confirmed rocky exoplanets from our sample is computed using equations~(\ref{eq:magmom} - \ref{eq:dynamo}) and plotted in Figure~\ref{fig:example_maxmm}.

Exoplanets with larger core radii have larger median ${\mathcal{M}}_{max}$ (Fig.~\ref{fig:example_maxmm}). As is evident by the location of the majority of the data, in the left hand side, the sample of currently detected exoplanets has an observational bias towards planets with short tidally-locked periods \citep{kipping2016observational}. In the future, a broader sample of exoplanet detections combined with better constrained rotation rate measurements, will contribute to the clarity of magnetic dipole moment trends. 

\begin{figure}
	\includegraphics[width=\columnwidth]{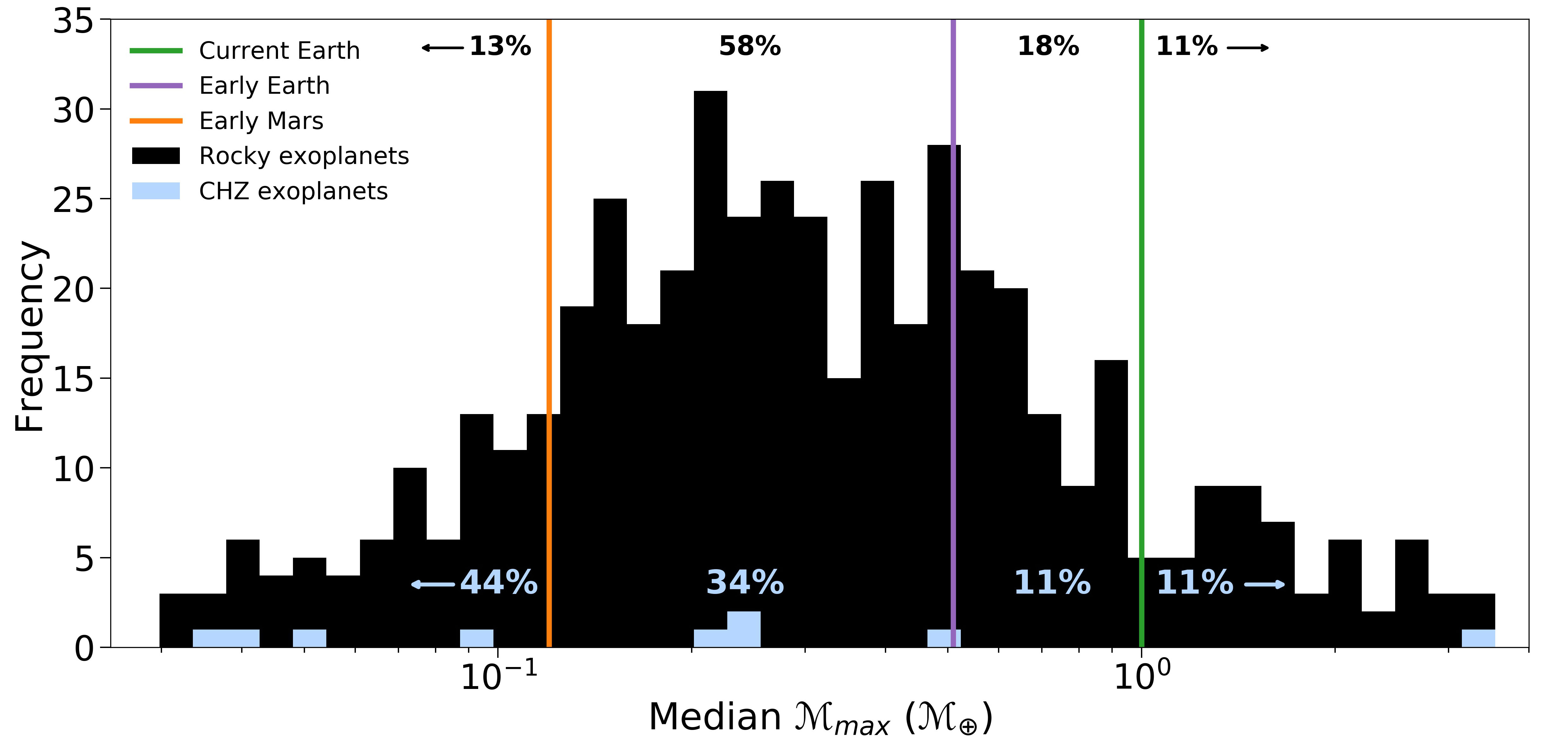}
    \caption{Histogram of the median ${\mathcal{M}}_{max}$ (from equations~(\ref{eq:magmom} - \ref{eq:dynamo})) of our 496 exoplanet sample (black) and the 9 CHZ subset (blue).}
    \label{fig:example_histgrm}
\end{figure}

\begin{figure}
	\includegraphics[width=\columnwidth]{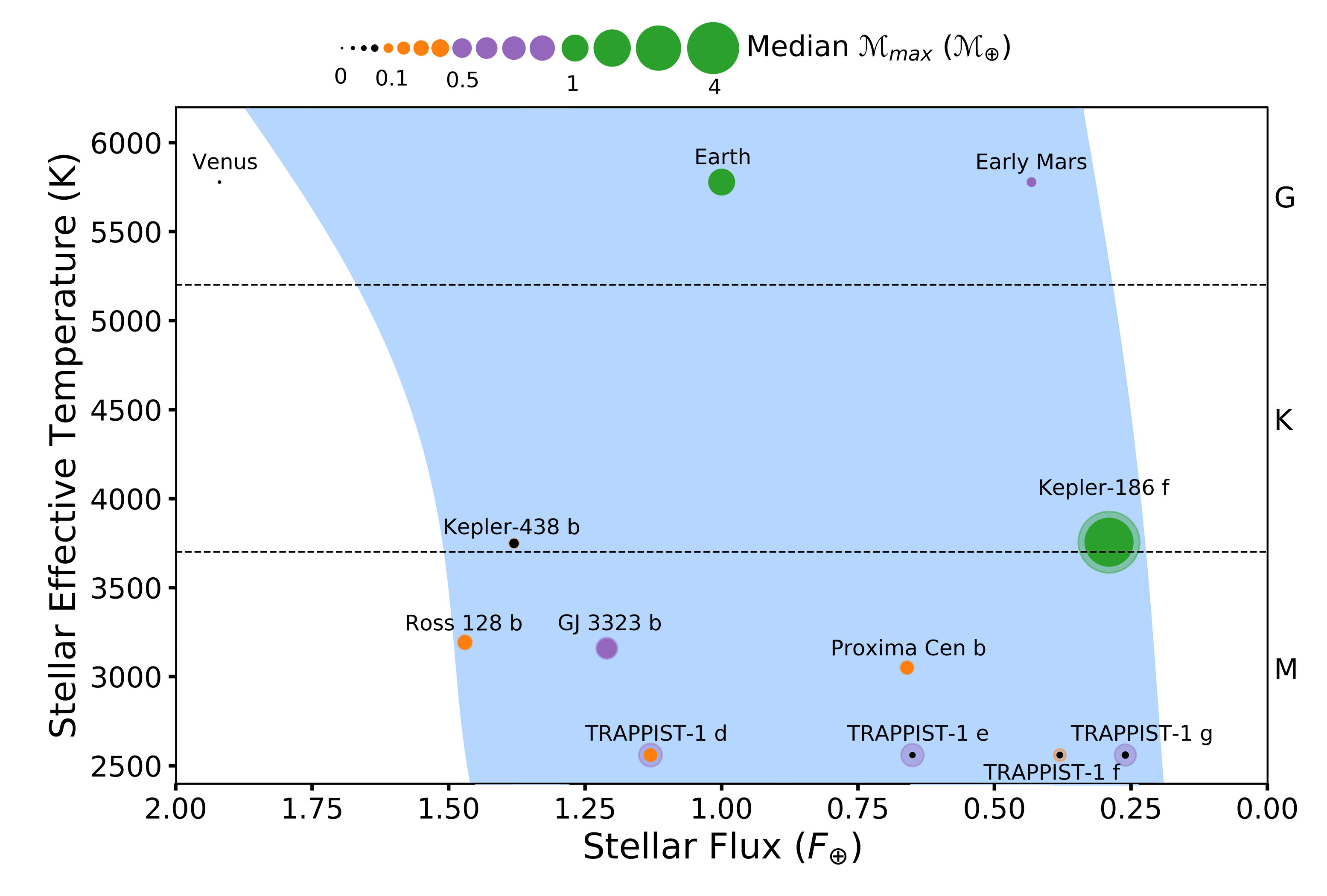}
    \caption{Potentially habitable rocky planets and their location within the CHZ \citep{kopparapu2013habitable,kopparapu2014habitable}. Point size indicates the planet's median  ${\mathcal{M}}_{max}$ values with upper 68\% confidence values displayed as lighter-coloured outer circles. Colours of the points indicate whether the exoplanet's ${\mathcal{M}}_{max} \geq 1{\mathcal{M}}_\oplus$ (green), $1{\mathcal{M}}_\oplus > {\mathcal{M}}_{max} \geq 0.5{\mathcal{M}}_\oplus$ (purple), $0.5{\mathcal{M}}_\oplus > {\mathcal{M}}_{max} \geq 0.1{\mathcal{M}}_\oplus$ (orange), or ${\mathcal{M}}_{max}<0.1{\mathcal{M}}_\oplus$ (black).}
    \label{fig:example_chz}
\end{figure}

Since there are no well-established connections between magnetic moment strength and habitability, to help place our modelled ${\mathcal{M}}_{max}$ values into a broader context, we compare our values to the values of three solar system scenarios involving the presence of liquid water on the surface of a planet - current Earth, early Earth and early Mars (Fig.~\ref{fig:example_histgrm} and Fig.~\ref{fig:example_chz}).

\citet{olson2006dipole} calculated Earth's current magnetic dipole moment as ${\mathcal{M}}_\oplus$ = 7.4 x 10${}^{22}$ A m${}^{2}$ and estimated the magnetic moment of early Mars: ${\mathcal{M}}_{\mars}$ = 0.9 x 10${}^{22}$ A m${}^{2}$ $\approx 0.1{\mathcal{M}}_{\oplus }$.  While the Earth's geodynamo appears to have been continuous since its inception, palaeomagnetic records suggest an increase in Earth's dynamo over the past  4 billion years \citep{tarduno2017earth}. Based on single silicate crystals and models of mantle cooling, the paleointensity data suggests an early Earth ($\sim$3.45 billion years ago) dipole moment of 3.8 x 10${}^{22}$ A m${}^{2}$ $\approx 0.51{\mathcal{M}}_{\oplus }$ \citep{tarduno2010geodynamo}. The reduced standoff of the solar wind during this period increases the potential for some atmospheric loss \citep{tarduno2010geodynamo,tarduno2007geomagnetic}. 

Figure~\ref{fig:example_histgrm} presents a frequency distribution of median ${\mathcal{M}}_{max}$ values for all exoplanets within our sample.

Taking into account the full Monte Carlo simulation described in section 2.3, in  our sample, 11\% $\pm$ 1\% (Fig.~\ref{fig:example_histgrm}) of all modelled exoplanets have median ${\mathcal{M}}_{max}$ larger than that of current Earth. This subset of exoplanets, with larger magnetic moments, would have better protection from the harmful effect of stellar and cosmic irradiation. Additionally, Fig.~\ref{fig:example_histgrm} shows that 13\% $\pm$ 1\% of all modelled exoplanets have median ${\mathcal{M}}_{max}$ less than early Mars.

Figure~\ref{fig:example_chz} displays the median ${\mathcal{M}}_{max}$ strength and upper 68\% confidence value (lighter-coloured outer circles) in the context of the CHZ.
\vskip 0.2in

In Fig.~\ref{fig:example_chz}, only Kepler-186 f has a median ${\mathcal{M}}_{max}$ larger than Earth's current dipole moment. With a large magnetic moment and temperate orbit, this planet would be the most likely to maintain the presence of surface liquid water over prolonged periods of time.

Figure~\ref{fig:example_chz} also shows that only GJ 3323 b has median ${\mathcal{M}}_{max}$ equal to or larger than that of early Earth, but smaller than the current Earth. Upon taking the 68\% upper limits into account additional planets Trappist-1 d, e and g, would fall into this subgroup. While we know there was life and liquid water present on the surface of Earth during the Paleoarchean, \citet{tarduno2010geodynamo,tarduno2007geomagnetic} discuss the potential for an increased rate of atmospheric and water loss during this period. If this lower magnetic dipole moment persisted over a prolonged period of time, an exoplanet would not be able to sustain a habitable environment due to the increased rate of water loss. Therefore, this subgroup of exoplanets would need a larger initial reservoir of water, or subsequent incoming supplies of water to reduce the effects of a lower magnetic dipole moment.

It has been hypothesised that Mars maintained liquid water during the period when it had a dynamo and magnetic moment value 0.1${\mathcal{M}}_\oplus$, however, it was unable to sustain the dynamo and experienced rapid loss of liquid water around the same time the dynamo went extinct \citep{ruiz2014early}. The 34\% $\pm$ 12\% (Fig.~\ref{fig:example_histgrm}) of rocky exoplanets from our CHZ subset with median ${\mathcal{M}}_{max}$ between that of early Mars and early Earth would not retain liquid water unless other circumstances were to counteract the effects of a lower magnetic moment, such as significant initial reservoir of water, high surface gravity, and/or a low amount of incident stellar irradiation.

Since a small magnetic moment has been connected with a loss of atmosphere \citep{vidotto2013effects}, depletion of water supplies \citep{elkins2013makes}, and lack of plate tectonics \citep{lammer2006loss}, the 44\% $\pm$ 13\% (Fig.~\ref{fig:example_histgrm}) of CHZ planets with median ${\mathcal{M}}_{max}<0.1{\mathcal{M}}_\oplus$, are unlikely to retain surface liquid water over prolonged periods of time. Thus, our results suggest that the CHZ might not be synonymous with the "liquid water zone" \citep{kaltenegger2017characterize}. 

Water is lost through a number of non-thermal processes such as dissociation and dissociative recombination \citep{geppert2008dissociative}, ion pickup \citep{lammer2006loss}, charge exchange \citep{dong2017dehydration}, sputtering \citep{terada2009atmosphere}, and atmospheric escape \citep{zahnle2017cosmic}. Future research should be conducted to more carefully investigate which water-loss processes dominate in different planetary scenarios. This would include an investigation of predicted atmospheric structures and the magnetopause positions relative to the planetary radius and atmospheric scale height.

\section{Conclusion}

Our analysis suggests that even with the best-case scenario of modelling the maximum magnetic dipole moment, 44\% $\pm$ 13\% of the currently detected rocky exoplanets in our sample, located in the CHZ have a negligible magnetic dipole moment. Since a lack of magnetic moment has been connected with the loss of atmosphere and depletion of water supplies, this group of exoplanets would not have protection against stellar and cosmic irradiation to maintain liquid water over a prolonged period of time, despite residing in the CHZ. Therefore, the habitability of each rocky exoplanet should be assessed on a case by case basis to determine which other factors apart from temperature and surface pressure could increase or decrease the chances of maintaining liquid water and hosting life. Planetary magnetism is one parameter amongst many astrophysical and geophysical properties of exoplanets and their host stars that could affect the habitability of a planet. These parameters need to be explored and evaluated in order to select the best targets for future observations characterising potentially habitable exoplanets. 

While the models used here provide a broad overview of planetary magnetism, it is important to recognise that our model represents the optimal case, and it is likely that many planets in our sample could have a magnetic moment significantly less than ${\mathcal{M}}_{max}$. Furthermore, the convective buoyancy flux equation does not account for variation or evolutionary changes to the internal structure of the planet and its core. Additionally, we have assumed that the rotation rates of our non-tidally locked planets (6/496) can be reasonably represented by the rotation rates of non-tidally locked solar system objects. Furthermore, 99\% of our sample is tidally locked, and therefore, the sample of rocky planets from transit surveys is heavily biased against habitable planets with strong magnetic dipole moments. Due to the slow rotations of tidally locked planets, according to our equations, they will have about 1/3 the magnetic field strength of non-tidally locked planets. The more we know about planetary and stellar structures and relations over time, the more accurate future models of planetary magnetism and habitability will become. Exoplanet detections from the TESS and PLATO missions, combined with ground-based radial velocities from existing and new spectrographs, will increase the sample size and improve the method used here.

\section{Acknowledgements}

We thank Mark Krumholz for helpful discussions and correspondence. S.R.N.McIntyre gratefully acknowledges an Australian Government Research Training Program (RTP)
Scholarship.




\bibliographystyle{mnras}
\bibliography{plmag}



\appendix
\section{Data Tables}

\begin{table*}
\caption{Rocky Exoplanets Sample (in order of decreasing dipole moment)}
\label{tab:exodat}
\begin{threeparttable}
\begin{tabular}{lccccc}
\hline
Planet Name & Mass (M$_\oplus$) & Radius (R$_\oplus$) & Orbital Period (days) & Stellar Age (Gyr) & Median ${\mathcal{M}}_{max}$ $({\mathcal{M}}_\oplus)$ \\
\hline
Kepler-186 f\tnote{$\ast$} \tnote{$\wedge$} & 1.76 $\pm$ 0.44 & 1.17 $\pm$ 0.08 & 129.944 $\pm$ 0.001 & 4.0 & 3.55 $\pm$ 2.40 \\
K04002.01 & 2.04 $\pm$ 0.44 & 1.22 $\pm$ 0.07 & 0.5241761 $\pm$ 0.0000007 & 1.4 & 3.21 $\pm$ 1.37 \\
K00284.04\tnote{$\ast$} & 1.53 $\pm$ 0.57 & 1.13 $\pm$ 0.11 & 110.2869 $\pm$ 0.0009 & 6.5 & 3.19 $\pm$ 2.24 \\
K02737.01 & 1.96 $\pm$ 0.29 & 1.21 $\pm$ 0.05 & 0.5266156 $\pm$ 0.0000007 & 6.2 & 3.10 $\pm$ 1.27 \\
K02874.01 & 1.22 $\pm$ 0.20 & 1.06 $\pm$ 0.05 & 0.3525160 $\pm$ 0.0000005 & 5.9 & 2.98 $\pm$ 1.11 \\
K2-157 b & 1.21 $\pm$ 0.53 & 1.06 $\pm$ 0.12 & 0.36526 $\pm$ 0.00003 & 1.0 & 2.88 $\pm$ 1.47 \\
K02916.01 & 0.94 $\pm$ 0.24 & 0.98 $\pm$ 0.07 & 0.3069375 $\pm$ 0.0000003 & 5.8 & 2.72 $\pm$ 1.08 \\
\hline
   \end{tabular}
   \end{threeparttable}
\end{table*}

\begin{table*}
\contcaption{Rocky Exoplanets Sample (in order of decreasing dipole moment)}
\label{tab:continued}
\begin{threeparttable}
\begin{tabular}{lccccc}
\hline
Planet Name & Mass (M$_\oplus$) & Radius (R$_\oplus$) & Orbital Period (days) & Stellar Age (Gyr) & Median ${\mathcal{M}}_{max}$ $({\mathcal{M}}_\oplus)$ \\
\hline
K01557.04 & 1.16 $\pm$ 0.19 & 1.04 $\pm$ 0.05 & 1.499147 $\pm$ 0.000003 & 1.0 & 0.93 $\pm$ 0.34 \\
K02675.02 & 0.86 $\pm$ 0.15 & 0.96 $\pm$ 0.05 & 1.116129 $\pm$ 0.000003 & 0.9 & 0.93 $\pm$ 0.34 \\
Kepler-847 b & 1.81 $\pm$ 0.39 & 1.18 $\pm$ 0.07 & 2.343232 $\pm$ 0.000008 & 3.9 & 0.92 $\pm$ 0.39 \\
K00569.02 & 1.22 $\pm$ 0.18 & 1.06 $\pm$ 0.04 & 1.632919 $\pm$ 0.000003 & 4.5 & 0.90 $\pm$ 0.33 \\
K02281.01 & 0.57 $\pm$ 0.33 & 0.85 $\pm$ 0.13 & 0.769854 $\pm$ 0.000001 & 1.7 & 0.90 $\pm$ 0.52 \\
K02781.01 & 1.20 $\pm$ 0.23 & 1.05 $\pm$ 0.06 & 1.615748 $\pm$ 0.000008 & 12.3 & 0.90 $\pm$ 0.34 \\
Kepler-1430 b & 1.81 $\pm$ 0.48 & 1.18 $\pm$ 0.09 & 2.46051 $\pm$ 0.00001 & 2.1 & 0.88 $\pm$ 0.39 \\
\hline
   \end{tabular}
   \end{threeparttable}
\end{table*}

\begin{table*}
\contcaption{Rocky Exoplanets Sample (in order of decreasing dipole moment)}
\label{tab:continued1}
\begin{threeparttable}
\begin{tabular}{lccccc}
\hline
Planet Name & Mass (M$_\oplus$) & Radius (R$_\oplus$) & Orbital Period (days) & Stellar Age (Gyr) & Median ${\mathcal{M}}_{max}$ $({\mathcal{M}}_\oplus)$ \\
\hline
K04382.01 & 1.22 $\pm$ 0.26 & 1.06 $\pm$ 0.06 & 2.94654 $\pm$ 0.00003 & 5.0 & 0.57 $\pm$ 0.22 \\
Kepler-746 b & 1.45 $\pm$ 0.54 & 1.11 $\pm$ 0.11 & 3.48159 $\pm$ 0.00001 & 8.7 & 0.57 $\pm$ 0.27 \\
K00717.02 & 0.37 $\pm$ 0.08 & 0.76 $\pm$ 0.05 & 0.900376 $\pm$ 0.000004 & 7.4 & 0.57 $\pm$ 0.25 \\
K03500.02 & 2.00 $\pm$ 0.50 & 1.21 $\pm$ 0.08 & 4.74841 $\pm$ 0.00008 & 2.9 & 0.57 $\pm$ 0.26 \\
Kepler-249 b & 1.36 $\pm$ 0.22 & 1.09 $\pm$ 0.05 & 3.30654 $\pm$ 0.00001 & 3.1 & 0.57 $\pm$ 0.21 \\
K02608.01 & 1.12 $\pm$ 0.31 & 1.03 $\pm$ 0.08 & 2.76122 $\pm$ 0.00001 & 1.0 & 0.56 $\pm$ 0.24 \\
K04298.01 & 1.14 $\pm$ 0.24 & 1.04 $\pm$ 0.06 & 2.89495 $\pm$ 0.00002 & 7.2 & 0.55 $\pm$ 0.21 \\
\hline
   \end{tabular}
   \end{threeparttable}
\end{table*}

\begin{table*}
\contcaption{Rocky Exoplanets Sample (in order of decreasing dipole moment)}
\label{tab:continued2}
\begin{threeparttable}
\begin{tabular}{lccccc}
\hline
Planet Name & Mass (M$_\oplus$) & Radius (R$_\oplus$) & Orbital Period (days) & Stellar Age (Gyr) & Median ${\mathcal{M}}_{max}$ $({\mathcal{M}}_\oplus)$ \\
\hline
K2-279 b & 1.98 $\pm$ 0.87 & 1.21 $\pm$ 0.14 & 7.123 $\pm$ 0.002 & 1.0 & 0.41 $\pm$ 0.23 \\
K04614.01 & 1.78 $\pm$ 0.47 & 1.17 $\pm$ 0.08 & 6.45473 $\pm$ 0.00008 & 4.4 & 0.41 $\pm$ 0.18 \\
K2-239 b & 1.41 $\pm$ 0.47 & 1.10 $\pm$ 0.10 & 5.240 $\pm$ 0.001 & 1.0 & 0.41 $\pm$ 0.19 \\
K02720.01 & 1.78 $\pm$ 0.24 & 1.17 $\pm$ 0.04 & 6.57158 $\pm$ 0.00002 & 8.1 & 0.40 $\pm$ 0.16 \\
K01480.02 & 1.89 $\pm$ 1.90 & 1.20 $\pm$ 0.28 & 7.00442 $\pm$ 0.00007 & 6.3 & 0.40 $\pm$ 0.39 \\
K04469.01 & 0.24 $\pm$ 0.06 & 0.67 $\pm$ 0.05 & 0.894172 $\pm$ 0.000002 & 2.5 & 0.40 $\pm$ 0.20 \\
K02208.01 & 0.62 $\pm$ 0.10 & 0.87 $\pm$ 0.04 & 2.344512 $\pm$ 0.000005 & 5.6 & 0.40 $\pm$ 0.15 \\
\hline
   \end{tabular}
   \end{threeparttable}
\end{table*}

\begin{table*}
\contcaption{Rocky Exoplanets Sample (in order of decreasing dipole moment)}
\label{tab:continued3}
\begin{threeparttable}
\begin{tabular}{lccccc}
\hline
Planet Name & Mass (M$_\oplus$) & Radius (R$_\oplus$) & Orbital Period (days) & Stellar Age (Gyr) & Median ${\mathcal{M}}_{max}$ $({\mathcal{M}}_\oplus)$ \\
\hline
K00284.03 & 1.12 $\pm$ 0.41 & 1.03 $\pm$ 0.10 & 6.17828 $\pm$ 0.00004 & 6.5 & 0.30 $\pm$ 0.14 \\
Kepler-119 c & 0.74 $\pm$ 1.53 & 0.92 $\pm$ 0.35 & 4.12510 $\pm$ 0.00003 & 3.6 & 0.30 $\pm$ 0.44 \\
K04032.03 & 1.07 $\pm$ 0.17 & 1.02 $\pm$ 0.05 & 5.99274 $\pm$ 0.00002 & 12.3 & 0.30 $\pm$ 0.11 \\
K01944.02 & 1.48 $\pm$ 0.24 & 1.12 $\pm$ 0.05 & 8.36060 $\pm$ 0.00008 & 0.5 & 0.29 $\pm$ 0.11 \\
K02722.04 & 1.55 $\pm$ 0.22 & 1.13 $\pm$ 0.04 & 8.92116 $\pm$ 0.00004 & 2.5 & 0.29 $\pm$ 0.11 \\
K00117.04 & 1.37 $\pm$ 0.24 & 1.09 $\pm$ 0.05 & 7.95830 $\pm$ 0.00006 & 4.3 & 0.29 $\pm$ 0.11 \\
K00671.04 & 1.93 $\pm$ 0.33 & 1.20 $\pm$ 0.06 & 11.13175 $\pm$ 0.00008 & 4.7 & 0.29 $\pm$ 0.12 \\
\hline
   \end{tabular}
   \end{threeparttable}
\end{table*}

\begin{table*}
\contcaption{Rocky Exoplanets Sample (in order of decreasing dipole moment)}
\label{tab:continued4}
\begin{threeparttable}
\begin{tabular}{lccccc}
\hline
Planet Name & Mass (M$_\oplus$) & Radius (R$_\oplus$) & Orbital Period (days) & Stellar Age (Gyr) & Median ${\mathcal{M}}_{max}$ $({\mathcal{M}}_\oplus)$ \\
\hline
K02859.04 & 0.36 $\pm$ 0.08 & 0.75 $\pm$ 0.05 & 2.90513 $\pm$ 0.00002 & 13.2 & 0.22 $\pm$ 0.10 \\
Kepler-1223 b & 2.04 $\pm$ 0.42 & 1.22 $\pm$ 0.07 & 16.3013 $\pm$ 0.0001 & 3.6 & 0.22 $\pm$ 0.10 \\
K01151.02 & 0.90 $\pm$ 0.14 & 0.97 $\pm$ 0.04 & 7.41086 $\pm$ 0.00003 & 8.5 & 0.22 $\pm$ 0.08 \\
K2-136 b & 0.96 $\pm$ 0.18 & 0.99 $\pm$ 0.05 & 7.9753 $\pm$ 0.0008 & 0.7 & 0.22 $\pm$ 0.08 \\
K03681.02 & 1.28 $\pm$ 0.19 & 1.07 $\pm$ 0.04 & 10.51420 $\pm$ 0.00005 & 2.3 & 0.22 $\pm$ 0.08 \\
K02585.02 & 1.26 $\pm$ 0.27 & 1.07 $\pm$ 0.06 & 10.42312 $\pm$ 0.00007 & 10.7 & 0.22 $\pm$ 0.09 \\
Kepler-1406 b & 1.41 $\pm$ 0.78 & 1.10 $\pm$ 0.16 & 11.62906 $\pm$ 0.00007 & 5.0 & 0.22 $\pm$ 0.13 \\
\hline
   \end{tabular}
   \end{threeparttable}
\end{table*}

\begin{table*}
\contcaption{Rocky Exoplanets Sample (in order of decreasing dipole moment)}
\label{tab:continued5}
\begin{threeparttable}
\begin{tabular}{lccccc}
\hline
Planet Name & Mass (M$_\oplus$) & Radius (R$_\oplus$) & Orbital Period (days) & Stellar Age (Gyr) & Median ${\mathcal{M}}_{max}$ $({\mathcal{M}}_\oplus)$ \\
\hline
K02961.01 & 0.31 $\pm$ 0.06 & 0.72 $\pm$ 0.04 & 3.78471 $\pm$ 0.00003 & 2.7 & 0.16 $\pm$ 0.07 \\
K04246.01 & 0.57 $\pm$ 0.14 & 0.85 $\pm$ 0.06 & 6.98472 $\pm$ 0.00006 & 0.9 & 0.16 $\pm$ 0.07 \\
K00246.02 & 0.76 $\pm$ 0.10 & 0.93 $\pm$ 0.03 & 9.60504 $\pm$ 0.00003 & 7.4 & 0.16 $\pm$ 0.06 \\
K02864.01 & 0.30 $\pm$ 0.05 & 0.72 $\pm$ 0.03 & 3.77887 $\pm$ 0.00004 & 8.3 & 0.16 $\pm$ 0.07 \\
K02722.05 & 1.32 $\pm$ 0.24 & 1.08 $\pm$ 0.05 & 16.5339 $\pm$ 0.0001 & 2.5 & 0.16 $\pm$ 0.06 \\
Kepler-1347 b & 1.11 $\pm$ 0.22 & 1.03 $\pm$ 0.06 & 14.00948 $\pm$ 0.00007 & 5.6 & 0.16 $\pm$ 0.06 \\
Kepler-1470 b & 1.27 $\pm$ 0.37 & 1.07 $\pm$ 0.09 & 16.2951 $\pm$ 0.0002 & 4.6 & 0.16 $\pm$ 0.07 \\
\hline
   \end{tabular}
   \end{threeparttable}
\end{table*}

\begin{table*}
\contcaption{Rocky Exoplanets Sample (in order of decreasing dipole moment)}
\label{tab:continued6}
\begin{threeparttable}
\begin{tabular}{lccccc}
\hline
Planet Name & Mass (M$_\oplus$) & Radius (R$_\oplus$) & Orbital Period (days) & Stellar Age (Gyr) & Median ${\mathcal{M}}_{max}$ $({\mathcal{M}}_\oplus)$ \\
\hline
K04725.01 & 0.34 $\pm$ 0.08 & 0.74 $\pm$ 0.05 & 7.28170 $\pm$ 0.00003 & 6.9 & 0.11 $\pm$ 0.05 \\
K02169.03 & 0.20 $\pm$ 0.03 & 0.64 $\pm$ 0.03 & 4.27225 $\pm$ 0.00002 & 1.9 & 0.10 $\pm$ 0.05 \\
K03436.01 & 2.09 $\pm$ 1.28 & 1.23 $\pm$ 0.19 & 44.2008 $\pm$ 0.0007 & 2.3 & 0.10 $\pm$ 0.07 \\
K03083.02 & 0.29 $\pm$ 0.07 & 0.71 $\pm$ 0.05 & 6.23219 $\pm$ 0.00006 & 2.5 & 0.10 $\pm$ 0.05 \\
K01151.03 & 0.24 $\pm$ 0.05 & 0.67 $\pm$ 0.04 & 5.24973 $\pm$ 0.00004 & 8.5 & 0.10 $\pm$ 0.05 \\
K01905.03 & 1.53 $\pm$ 0.28 & 1.13 $\pm$ 0.06 & 34.2113 $\pm$ 0.0003 & 3.2 & 0.10 $\pm$ 0.04 \\
Kepler-388 c & 0.58 $\pm$ 0.28 & 0.86 $\pm$ 0.11 & 13.2970 $\pm$ 0.0002 & 4.0 & 0.10 $\pm$ 0.05 \\
\hline
   \end{tabular}
   \end{threeparttable}
\end{table*}

\begin{table*}
\contcaption{Rocky Exoplanets Sample (in order of decreasing dipole moment)}
\label{tab:continued7}
\begin{threeparttable}
\begin{tabular}{lccccc}
\hline
Planet Name & Mass (M$_\oplus$) & Radius (R$_\oplus$) & Orbital Period (days) & Stellar Age (Gyr) & Median ${\mathcal{M}}_{max}$ $({\mathcal{M}}_\oplus)$ \\
\hline
K00082.03 & 0.36 $\pm$ 0.06 & 0.75 $\pm$ 0.03 & 27.4537 $\pm$ 0.0001 & 1.1 & 0.04 $\pm$ 0.02 \\
K03184.02 & 0.15 $\pm$ 0.04 & 0.59 $\pm$ 0.05 & 11.3220 $\pm$ 0.0006 & 4.4 & 0.04 $\pm$ 0.02 \\
K00082.05 & 0.07 $\pm$ 0.01 & 0.47 $\pm$ 0.02 & 5.28695 $\pm$ 0.00003 & 1.1 & 0.04 $\pm$ 0.02 \\
TRAPPIST-1 e\tnote{$\wedge$} & 0.62 $\pm$ 0.58 & 0.92 $\pm$ 0.04 & 6.09962 $\pm$ 0.00001 & 0.5 & 0.03 $\pm$ 0.78 \\
K01616.02 & 0.38 $\pm$ 0.36 & 0.76 $\pm$ 0.17 & 34.775 $\pm$ 0.001 & 2.9 & 0.03 $\pm$ 0.03 \\
Kepler-444 b & 0.04 $\pm$ 0.01 & 0.40 $\pm$ 0.02 & 3.600105 $\pm$ 0.000008 & 11.2 & 0.03 $\pm$ 0.02 \\
Kepler-378 c & 0.28 $\pm$ 0.07 & 0.70 $\pm$ 0.05 & 28.9060 $\pm$ 0.0002 & 3.2 & 0.03 $\pm$ 0.01 \\
K01961.02 & 0.71 $\pm$ 0.25 & 0.91 $\pm$ 0.08 & 76.642 $\pm$ 0.002 & 13.2 & 0.03 $\pm$ 0.01 \\

   \hline
   \end{tabular}
  \begin{tablenotes}
   \item[$\ast$] Non-tidally locked planets
   \item[$\wedge$] Planets located in the optimistic CHZ as defined by \citep{kopparapu2013habitable,kopparapu2014habitable}.
   \end{tablenotes}
   \end{threeparttable}
\end{table*}

\begin{table*}
\caption{Rotation rate}
\label{tab:rotcal}
\begin{threeparttable}
\begin{tabular}{llccl}
\hline
Object Name & Object Type & Rotation Period (days) & Rotation Rate ($\Omega_\oplus$) & Reference\\
\hline
Mercury & planet & 58.646 & 0.017 & Williams 2017\tnote{4}\\
Venus & planet & 243.019 & 0.004 & Williams 2017\tnote{4}\\
Earth & planet & 1.000 & 1.000 & Williams 2017\tnote{4}\\
Mars & planet & 1.026 & 0.975 & Williams 2017\tnote{4}\\
16 Psyche & asteroid & 0.175 & 5.716 & \citet{sheppard2002time}\\
107 Camilla & asteroid & 0.200 & 5.002 & \citet{sheppard2002time}\\
87 Sylvia & asteroid & 0.217 & 4.617 & \citet{sheppard2002time}\\
\hline
\end{tabular}
\begin{tablenotes}
     \item[2] {\url{https://nssdc.gsfc.nasa.gov/planetary/factsheet/saturniansatfact.html}}
     \item[3] {\url{https://nssdc.gsfc.nasa.gov/planetary/factsheet/joviansatfact.html}}
     \item[4] {\url{https://nssdc.gsfc.nasa.gov/planetary/factsheet/}}
   \end{tablenotes}
   \end{threeparttable}
\end{table*}


\bsp	
\label{lastpage}
\end{document}